\documentclass[5p]{elsarticle}
\usepackage{graphicx}
\usepackage{amsmath,amsfonts,amssymb}
\usepackage{xfrac}
\usepackage{bm}
\usepackage{algorithm}
\usepackage{algorithmic}
\usepackage{afterpage}
\usepackage{url}
\usepackage[breaklinks=true,letterpaper=true,colorlinks,bookmarks=false]{hyperref}
\usepackage[table]{xcolor}
\usepackage{booktabs}  % For toprule
\usepackage{multirow}
\usepackage{multicol}
\usepackage[switch]{lineno}
\modulolinenumbers[5]

\biboptions{authoryear}

\definecolor{deep_blue}{rgb}{0,.2,.5}
\definecolor{dark_blue}{rgb}{0,.15,.5}

\hypersetup{pdftex,  % needed for pdflatex
  breaklinks=true,  % so long urls are correctly broken across lines
  colorlinks=true,
  linkcolor=dark_blue,
  citecolor=deep_blue,
}

\def\slantfrac#1#2{\kern.1em^{#1}\kern-.1em/\kern-.1em_{#2}}
\newif\iffinal
\finaltrue
\iffinal\else
\usepackage[tightpage,active]{preview}
\fi

\usepackage{xcolor}
\usepackage[normalem]{ulem}
\colorlet{markercolor}{purple!50!black}

\usepackage{MnSymbol}
\begin{document}
% Show line numbers
%\linenumbers
%\linenumbersep 3pt\relax
%\renewcommand\linenumberfont{\normalfont\tiny\sffamily\color{black!50}}

\journal{NeuroImage}
\title{Cross-validation failure: small sample sizes lead to large error bars}

\author[parietal,cea]{Ga\"el Varoquaux\corref{corresponding}}

\cortext[corresponding]{Corresponding author}

\address[parietal]{Parietal project-team, INRIA Saclay-\^ile de France,
France}
\address[cea]{CEA/Neurospin b\^at 145, 91191 Gif-Sur-Yvette, France}

\begin{abstract}
Predictive models ground many state-of-the-art developments in statistical brain image analysis: decoding, MVPA, searchlight, or extraction of biomarkers. The principled approach to establish their validity and usefulness is cross-validation, testing prediction on unseen data. Here, I would like to raise awareness on error bars of cross-validation, which are often underestimated. Simple experiments show that sample sizes of many neuroimaging studies inherently lead to large error bars, \emph{eg} $\pm 10\%$ for 100 samples. The standard error across folds strongly underestimates them. These large error bars compromise the reliability of conclusions drawn with predictive models, such as biomarkers or methods developments where, unlike with cognitive neuroimaging MVPA approaches, more samples cannot be acquired by repeating the experiment across many subjects. Solutions to increase sample size must be investigated, tackling possible increases in heterogeneity of the data.
\end{abstract}


\maketitle%

%%%%%%%%%%%%%%%%%%%%%%%%%%%%%%%%%%%%%%%%%%%%%%%%%%%%%%%%%%%%%%%%%%%%%%%%%%%%%%%%
\smash{\raisebox{30em}{{\sffamily\bfseries Comments and Controversies}}}%
\vspace*{-3em}%

\sloppy % Fed up with messed-up line breaks

\section{Introduction}%

% Start positive
In the past 15 years, machine-learning methods have pushed forward many
brain-imaging problems: decoding the neural support of cognition
\citep{haynes2006},
information mapping \citep{kriegeskorte2006}, prediction of individual differences --behavioral or
clinical-- \citep{smith2015positive}, rich encoding models
\citep{nishimoto2011}, principled reverse inferences
\citep{poldrack2009decoding}, \emph{etc}. Replacing in-sample statistical testing by
prediction gives more power to fit rich
models and complex data \citep{norman2006,varoquaux2014machine}.

% Introduce cross-validation
The validity of these models is established by their ability to generalize:
to make accurate predictions about some properties of \emph{new} data.
They need to be tested
on data independent from the data used to fit them. Technically, this
test is done via \emph{cross-validation}: the available data is split in
two, a first part, the \emph{train set} used to fit the model, and a
second part, the \emph{test set} used to test the model
\citep{pereira2009machine,varoquaux2017assessing}.

% State the problem
Cross-validation is thus central to statistical control of the 
numerous neuroimaging techniques relying on machine learning: decoding,
MVPA (multi-voxel pattern analysis), searchlight, computer aided
diagnostic, \emph{etc}. \cite{varoquaux2017assessing} conducted a 
review of cross-validation techniques with an empirical study on
neuroimaging data. These experiments revealed that cross-validation made
errors in measuring prediction accuracy typically around $\pm 10\%$.
Such large error bars are worrying.

% Explain what I do, and our conclusion
Here, I show with very simple analyses that the observed errors of
cross-validation are inherent to small number of samples. I argue
that they provide loopholes that are exploited in the neuroimaging
literature, probably unwittingly. The problems are particularly severe
for methods development and inter-subject diagnostics studies.
Conversely, cognitive neuroscience studies are less impacted, as they often have access to higher
sample sizes using multiple trials per subjects and multiple subjects.
These issues could undermine the potential of machine-learning methods in
neuroimaging and the credibility of related publications. I give
recommendations on best practices and explore cost-effective avenues to
ensure reliable cross-validation results in neuroimaging.

The effects that I describe are related to the ``power failure'' of
\cite{button2013power}: lack of statistical power. In the specific case
of testing predictive models, the shortcoming of small samples are more
stringent and inherent as they are not offset with large effect sizes.
My goals here are to raise awareness that studies based on predictive modeling
require larger sample sizes than standard statistical approaches.

\section{Results: cross-validation errors}

\begin{figure*}[t]
    \setlength{\fboxsep}{0pt}%
    \begin{minipage}[T]{.26\paperwidth}
     \includegraphics[height=.28\paperwidth]{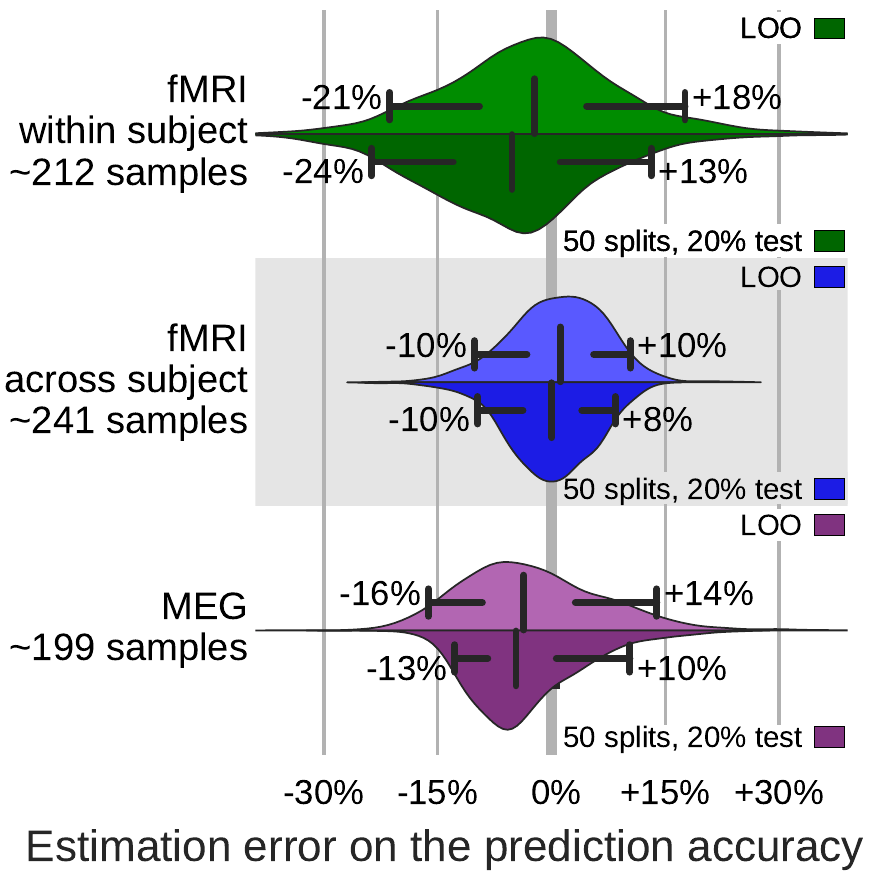}%
     \llap{\raisebox{.265\paperwidth}{%
	\parbox{.28\paperwidth}{\colorbox{white}{\bfseries\sffamily
	a\hspace*{-.15ex}.\,Neuroimaging data\hspace{-2ex}}}}}%
    \medskip

     \includegraphics[width=1.11\linewidth]{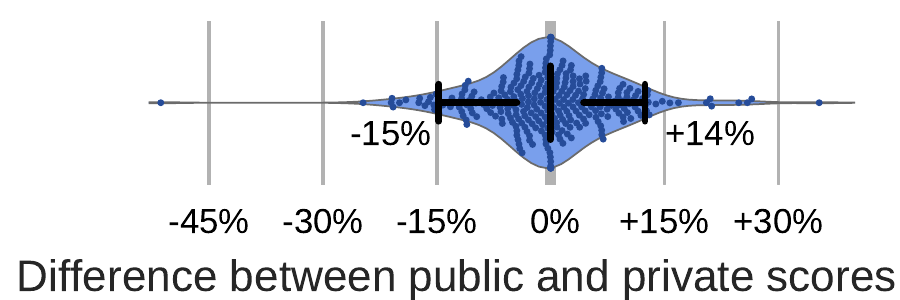}%
     \llap{\raisebox{.3\linewidth}{%
	\parbox{1.11\linewidth}{\colorbox{white}{%
	    \bfseries\sffamily d\hspace*{-.1ex}.\,Kaggle
	    competition\hspace*{-1ex}}}}}%
    \end{minipage}%
    \hfill%
    \begin{minipage}[T]{.25\paperwidth}\vspace*{-2.7ex}
     \includegraphics[height=.365\paperwidth]{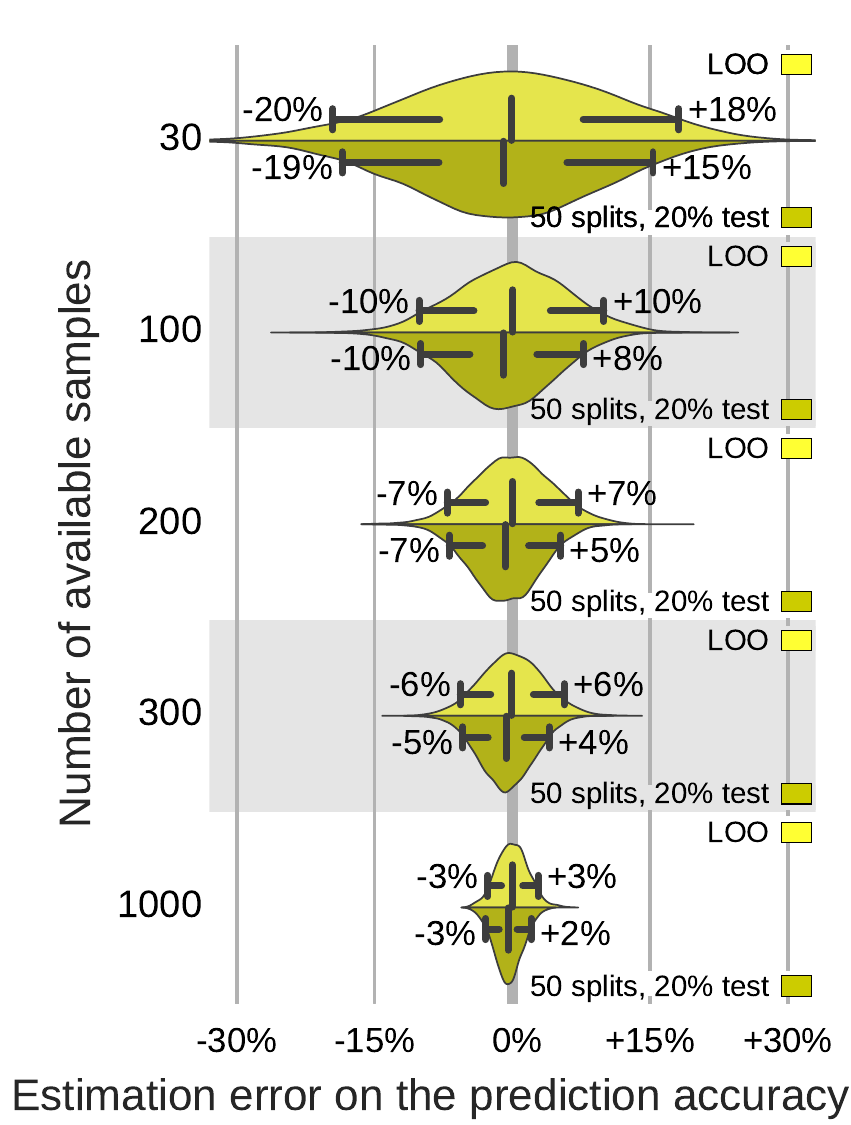}%
     \setlength{\fboxsep}{1pt}%
     \llap{\raisebox{.345\paperwidth}{%
	\parbox{.23\paperwidth}{\colorbox{white}{\bfseries\sffamily
	b\hspace*{-.13ex}.\,Simulations}}}}%
    \smallskip

    \end{minipage}%
    \hfill%
    \begin{minipage}[T]{.26\paperwidth}\vspace*{-2ex}
     \includegraphics[height=.36\paperwidth]{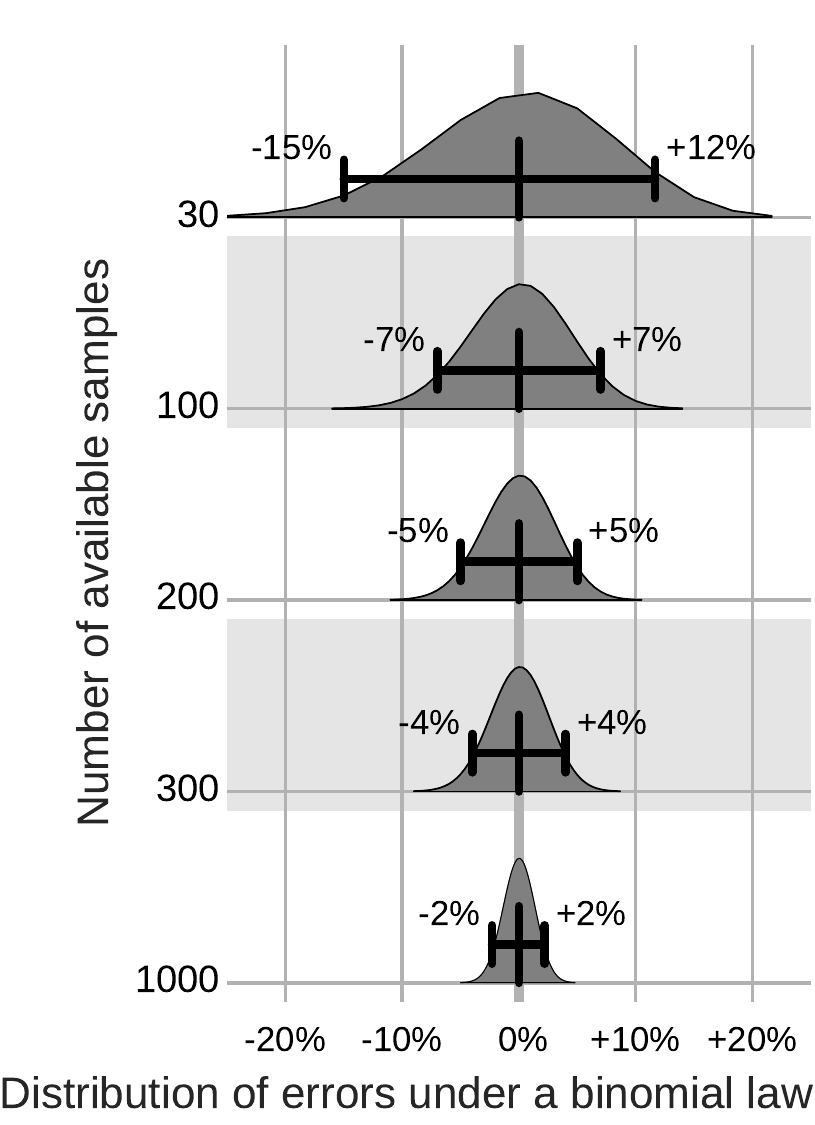}%
     \llap{\raisebox{.34\paperwidth}{%
	\parbox{.23\paperwidth}{\colorbox{white}{\bfseries\sffamily
	c\hspace*{-.11ex}.\,Binomial law}}}}%
    \smallskip

    \end{minipage}
    \caption{\textbf{Cross-validation errors.}
    {\bfseries\sffamily a} -- Distribution of errors between the prediction accuracy as
    assessed via cross-validation (average across folds) and as measured on a large independent test
    set for different types of neuroimaging data.
    Results from \cite{varoquaux2017assessing} (see
    \ref{sec:varoquaux_2017})
    {\bfseries\sffamily b} -- Distribution of errors between the prediction accuracy as assessed
    via cross-validation on data of various sample sizes and as measured
    on 10\,000 new data points for simple simulations (see
    \ref{sec:simulated_data}).
    {\bfseries\sffamily c} -- Distribution of errors as given by a binomial
    law: difference between the observed prediction error and the 
    population value of the error, $p = 75\%$,
    for different sample sizes.
    {\bfseries\sffamily d} -- Discrepancies between private and public
    score.
    Each dot represents the difference between the accuracy of a method on
    the public test data and on the private one. The scores are retrieved
    from
    \href{https://www.kaggle.com/c/mlsp-2014-mri}{www.kaggle.com/c/mlsp-2014-mri},
    in which 144
    subjects were used total, 86 for training the predictive model, 30 for 
    the public test set, and 28 for the private test set.
    The bar and whiskers indicate the median
    and the 5\raisebox{.5ex}{\tiny th} and 95\raisebox{.5ex}{\tiny th}
    percentile. 
    Measures on cross-validation (a and b) are reported for
    two reasonable choices of cross-validation strategy: leave one out
    (leave one run out or leave one subject out in data with multiple runs or
    subjects), or 50-times repeated splitting of 20\% of the data.
    \label{fig:cross_val_errors}}
\end{figure*}

\subsection{Distribution of errors in cross-validation}

Cross-validation strives to measure the generalization power of a model:
how well it will predict on new data. To simplify the discussion, I will
focus on balanced classification, predicting two categories of samples;
prediction accuracy can then be measured in percents and
chance is at 50\%. The cross-validation error
is the discrepancy between the prediction accuracy measured by
cross-validation and the expected accuracy on new data.

\paragraph{Previous results: cross-validation on brain images}

\cite{varoquaux2017assessing} used a nested cross-validation on
neuroimaging data to measure this discrepancy: we split the data multiple
times and compared errors (see \ref{sec:varoquaux_2017}). The strength of such an experiment is that it
is applied on actual neuroimaging data, mimicking usage by practitioners.
Its weakness is that the models' true generalization accuracy is not
known and must be estimated.

\autoref{fig:cross_val_errors}a summarizes the resulting
cross-validation errors, show a similar behavior across 
different reasonable choices of cross-validation strategy: the common
leave-one-run-out, and the recommended random splitting strategy
\citep{varoquaux2017assessing}.
The 5\raisebox{.5ex}{\small th} and
95\raisebox{.5ex}{\small th} percentile of the distribution of errors are
of particular interest as they correspond to the commonly accepted .05
threshold on p-values. The results show that these confidence bounds
extends at least 10\% \emph{both ways}, regardless of the
cross-validation strategy used.  It implies
that, when computing a given cross-validated accuracy, there is a 5\%
chance that it is 10\% above the true generalization accuracy, and a
5\% chance this it is 10\% below.

\paragraph{Spread out predictions in a public challenge}

There could be something unusual in the settings of
\cite{varoquaux2017assessing}. To reflect common practice in
neuroimaging, I have inspected the results of a public prediction
challenge \citep{silva2014tenth} on the Kaggle
website\footnote{\url{https://www.kaggle.com/c/mlsp-2014-mri}}. The competition
--predicting Schizophrenia diagnosis from functional and structural MRI--
reports two accuracy measures estimated on a public ($n=30$) and a
private ($n=28$) test set.

The accuracy scores reported on the public and the private test set show
a large difference. \autoref{fig:cross_val_errors}d summarizes these differences.
Computing confidence bounds from these discrepancies gives errors on the
order of $\pm 15\%$. As neither the public nor the private test set is a
gold standard, it is reasonable to assume that errors are shared between
the two scores, and thus the actual margin of error on a single
measurement is smaller by a factor of two.

\paragraph{Simple simulations also display large error bars}

To understand better the origin of these discrepancies, I used simple
simulations: fitting a linear SVM on a two-class dataset, samples drawn
\emph{i.i.d.} from two Gaussian distributions with a separation tuned
such that the classifier achieves 75\% accuracy. I then compare the
prediction accuracy measure by cross-validation on these data with the
accuracy that the classifier achieved on a large amount (10\,000) new
samples drawn from the same distribution. An important benefit of this
experiment is that it shows the difference between the cross-validation
measure of the classifier's accuracy, and the \emph{true} generalization
accuracy.

\autoref{fig:cross_val_errors}b shows the resulting distribution of errors on
the prediction accuracy estimated by cross validation for different size
of the data available. For 100 samples, these experiments reproduces well
the errors observed on neuroimaging data
(\autoref{fig:cross_val_errors}a and \ref{fig:cross_val_errors}d). 
Both leave one out and more sophisticated cross-validation strategies
display large error bars\footnote{Performing 50 repeated splits of 20\%
of the data yields slightly smaller error bars than leave one out, and
can be significantly less computationally expensive for large datasets. This
cross-validation strategy should be preferred, but will not fix the
problem of large error bars.}. As the sample size of the
simulated data goes up, the error bars narrow markedly.

\paragraph{Intrinsically large sampling noise}
The data clearly shows that the accuracy of predictive models is not well
measured in neuroimaging. The small sample sizes encountered in
neuroimaging indeed make this task very challenging: as I show below,
even in ideal situations, there is a large sampling noise in the measure.

\begin{figure}
    \centerline{%
	\includegraphics[width=\linewidth]{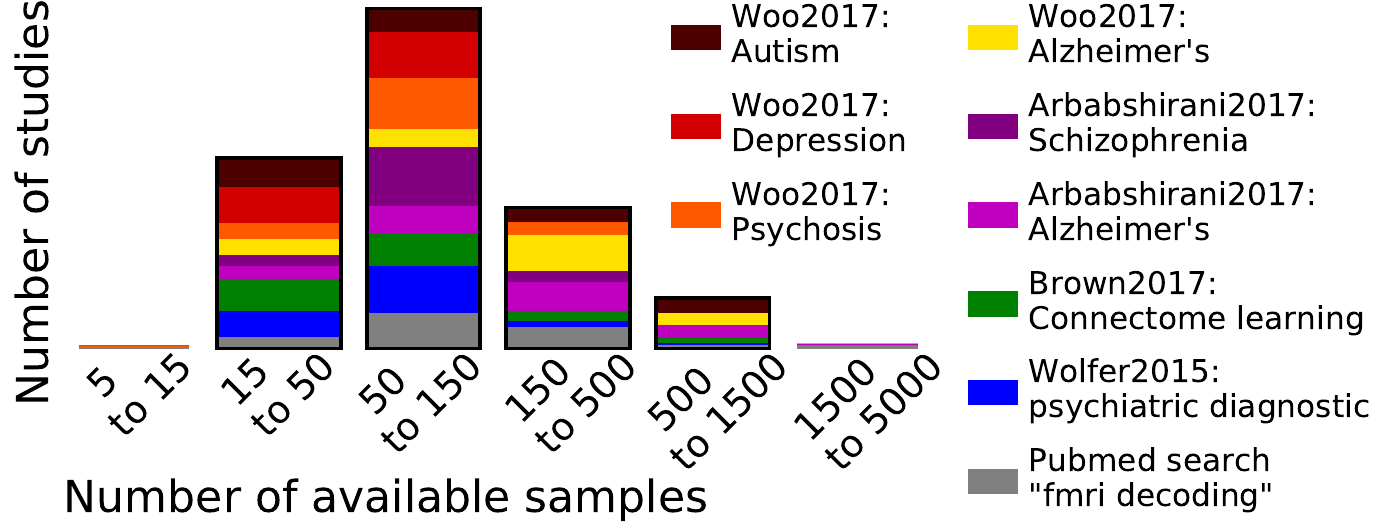}%
    }%
    \caption[Sample sizes in neuroimaging studies]{\textbf{Sample sizes in neuroimaging studies}
    A stacked histogram compounding sample sizes from multiple
    sources: the \cite{woo2017building} review paper, 
    differentiating Autism, Depression, Pyschosis, and Alzheimer's studies,
    the \cite{arbabshirani2017single} review paper, 
    differentiating Schizophrenia and Alzheimer's studies, the
    \cite{brown2016machine} review on prediction from connectomes, and
    the \cite{wolfers2015estimating} review on prediction for psychiatric
    disorders, as
    well as the 100 first answers to a
    pubmed search on "fmri decoding".
    The total histogram comprises 642 studies, with a median number of samples 
    of 89.

    Note that I did not consider groups of less than 25 studies, and
    hence did not break up into pathologies the \cite{brown2016machine}
    and \cite{wolfers2015estimating} reviews.
    \label{fig:sample_size}
    }
\end{figure}

The typical sample size of neuroimaging studies is less than 100
observations given to the classifier,
trials or subjects depending on the settings (\autoref{fig:sample_size}).
The simplest model for the observed prediction
errors is that of tossing a coin 100 times with a probability $p$
of success at each toss. The probability $p$ corresponds to the accuracy
of the classifier that we are trying to measure. The distribution of
number of successes is then given by a binomial law 
\citep{pereira2011information,stelzer2013statistical}. With 100 tosses,
associated confidence bounds lie $\pm 7\%$ away from the true
accuracy $p$ (see \autoref{fig:cross_val_errors}c).

% Cite Joe's paper that confirms that in practice it's worst than
% binomial

% Tell us why it's worst: correlated data and dimensionality
This binomial law is a best-case scenario for errors on the accuracy
measure: observations are \emph{i.i.d.}~and there is no additional
variability from training a decoder. On the opposite, neuroimaging
data is strife with correlation across samples and confounding effects,
\emph{e.g.}~the temporal structure of trials or samples drawn either from 
the same subject or different subjects.
These reduce the statistical degrees of freedom and create an
intrinsic variance in the prediction accuracy
\citep{saeb2017,little2017}. This is why we observe that cross-validation has
larger errors on
neuroimaging data (\autoref{fig:cross_val_errors}a) than on the
simulations (\autoref{fig:cross_val_errors}b) or with the ideal 
binomial law (\autoref{fig:cross_val_errors}c).

%\section{Distribution of errors under a binomial law}

%The simplest model for prediction errors in a two class problem is the
%binomial law: each observation has a constant probability $p$ of being
%classified correctly or incorrectly. \autoref{fig:binomial} shows the
%difference between the observed error and $p$ under such a distribution
%for different sample sizes and $p=75\%$.

%Distributions of errors in cross-validation will be wider than this
%distribution. Indeed, it is an ideal case in which all prediction errors
%are \emph{i.i.d.} --which requires more than $\emph{i.i.d.}$
%observations, as overlapping train or test folds can create dependencies
%in prediction errors. Compared to the ideal predictor, the error bounds
%are indeed slightly more narrow. Nevertheless, this sampling noise
%explains a significant part of the error in measuring prediction
%accuracy.

% Maybe here is a good idea to mention that it is a problem in the test
% set, and not the train set: better predictors do not improve the
% situation.

% At some point, I need to mention that things are better if errors are
% correlated when we compare two things, as in methods development.

% Explain why this is less of a problem in standard statistics: low
% dimensions

\bigskip

Simulations and a simple null model therefore show
that the error bars of cross-validation observed in neuroimaging are
perfectly expected given the sample sizes. Improvements on
cross-validation such as the reusable holdout \citep{dwork2015reusable}
cannot circumvent intrinsic limitations of small samples (see
\ref{sec:additional_considerations}).

\subsection{Small sample sizes undermine statistical control}

\paragraph{Underestimated errors}
Not only are the errors of cross-validation large, but it is also easy
to underestimate them, as when using as a null the binomial distribution.

The simplest approach to put error bars on cross-validation results is
to look at the dispersion of the prediction accuracy across the folds.
However as the predictions are not independent across folds, 
estimates of the variance or related statistical tests are
optimistic \citep{bengio2004no}. On the simulated data,
formulas based on the standard error to mean underestimate confidence
bounds by a factor of 0.7 in the best case
(\ref{sec:sem_error}).

\label{sec:underestimated}

% Check out page 78 of Joe's paper

Permutation testing gives good statistical control on the prediction
accuracy \citep{stelzer2013statistical}.
Literature search on Google Scholar\footnote{%
Pubmed does not do full-text search.
On Google scholar, a search for
\href{https://scholar.google.fr/scholar?q=fmri+decoding&hl=en&as_sdt=0\%2C5&as_ylo=2012&as_yhi=}{``fmri decoding''} in the last 5 years returned 
15\,500 results, while 
\href{https://scholar.google.fr/scholar?q=fmri+decoding+permutation&hl=en&as_sdt=0\%2C5&as_ylo=2012&as_yhi=}{``fmri
decoding permutation''} returned
2380; similarly, 
\href{https://scholar.google.fr/scholar?q=fmri+mvpa&btnG=&hl=en&as_sdt=0\%2C5&as_ylo=2012}{``fmri
mvpa''} return 2360 results while
\href{https://scholar.google.fr/scholar?q=fmri+mvpa+permutation&btnG=&hl=en&as_sdt=0\%2C5&as_ylo=2012}{``fmri
mvpa permutation''} return 728.%
} suggest that around a 30\% of the publications on MVPA (mostly
searchlight-based analysis) use permutations, but that only
15\% of the fMRI decoding studies use permutations.

% Compare to power estimates in standard statistics: compare the
% recommended sample sizes

\paragraph{Vibration effects}

Analytic pipelines come with various methodological choices that are hard to
settle a priori \citep{carp2012secret}. With a high-variance test
statistic, as cross validation on few samples, methodological
choices can have a drastic impact on the outcome of the analysis. This is
sometimes known as \emph{vibration}, and the key quantity is the ratio
between the effect size and the variations due to analytical choices
\citep{ioannidis2008most}. I explored vibration effects in decoding using
the face versus place opposition in the \cite{haxby2001} data. I
inverted the labels to predict in one session out of two, to create a
dataset in which fMRI should not predict the experimental condition. On
this data, I ran a variety of classic decoding pipelines, namely SVM or
logistic regression, optionally with feature selection of 100, 200, 500,
1\,000, or 2\,000 voxels and smoothing at 2, 4, or 6\,mm. These
are standard choices, but they give altogether almost 50 different
related decoding pipelines. I applied all these pipelines to various
subsets of the data: the full 12 sessions, the 6 first or 6 last, or the
4 first or 4 last sessions.

\label{sec:vibration}

\begin{figure}
    \centerline{%
	\includegraphics[width=.8\linewidth]{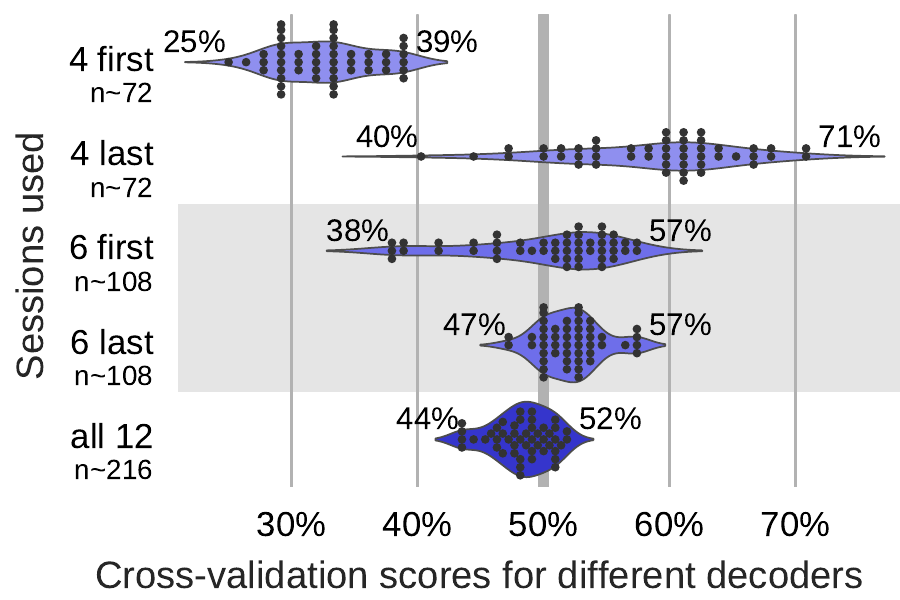}%
    }%
    \caption{\textbf{Different decoders on fMRI with permuted labels}
    On each line shows the distribution of cross-validation scores for a
    variety of decoders (SVC and logistic regression, with different
    amount of univariate feature selection and spatial smoothing);
    a dot is the cross-validation score for one choice of decoder. These
    are applied to the fMRI data of the first subject in
    \cite{haxby2001}, discriminating face viewing and place viewing, but
    with labels inverted one session out of two; hence the expected
    accuracy is chance: 50\%.
    \label{fig:haxby_decoding}
    }
\end{figure}

\autoref{fig:haxby_decoding} shows the cross-validation scores obtained
with the various pipelines. The expected prediction score is 50\%, chance.
When using all 12 sessions, the observed scores group well around 50\%,
with excursions ranging from 44\% to 52\%. However, when using less data
the excursions are much more pronounced, going up to 57\% for 6 sessions
and 71\% for 4 sessions. In addition, the mean observed score varies notably across
subsets of the data. Such variation can be explained by
nonstationarities, \emph{e.g.} fluctuation of attention of the subject,
or sampling noise discussed above: the observations are very correlated
and thus $n \sim 100$ may not represent well the faces and places
conditions.

\section{Implications for neuroimaging}

\subsection{An open door to overfit and confirmation bias}
% Implication on the field

The large error bars are worrying, whether it is for methods development
of predictive models or their use to study the brain and the mind.
Indeed, a large variance of results combined with publication incentives
weaken scientific progress \citep{ioannidis2005most}.

% Mention the link to overfit via researchers degrees of freedom
With conventional statistical hypothesis testing, the danger of vibration
effects is well recognized: arbitrary degrees of freedom in the analysis 
explore the variance of the results and, as a consequence,
control on false positives is easily lost 
\citep{simmons2011false}. 
% cite the carp paper
\citep{carp2012secret} has found that the variety of analytics choices is
such in fMRI that almost every publication uses a unique pipeline.
In predictive models,
arbitrary choices can leads to artificial improvements in the prediction
accuracy measured by cross-validation (see \autoref{sec:vibration} and
\cite{skocik2016tried}). The larger is the variance of the measure of the
prediction score, the larger are these effects. The
improvements are meaningless as they will not carry over to predicting on
new data. The danger is
well known in machine learning, where it is known as \emph{overfit}. The
standard remedy is to keep a large independent test set. However it is
difficult in neuroimaging, where data acquisition is costly.
To mitigate such intrinsic problems, clinical trials often use blind
analysis where part of the labels are unknown to the statistician.

% Read / cite
% Palatucci, M. and Carlson, A. (2008). On the chance accuracies of large collections of
% classifiers. In Proceedings of the 25th International Conference on Machine Learning

% Publication bias
Scientific publishing makes things worse: the literature acts as a filter
as only studies that report
significant effects are published. Such selective reporting can further
undermine control of the fraction of false detections in a body of
literature \citep{rosenthal1979file}. It also tends to inflate the
reported effect size \citep{vul2009puzzlingly}. 
An additional a dangerous effect of large variance is that it
enables and justifies confirmation bias in publications: investigators
or reviewers are more likely to publish results that are in agreement
with their theory. Analysis of the literature suggests that
publications are indeed too often on the edge of significance
\citep{szucs2016empirical} and are vastly biased
by selection according to the prevailing opinions \citep{ioannidis2008most}.

\begin{figure}
    \includegraphics[width=\linewidth]{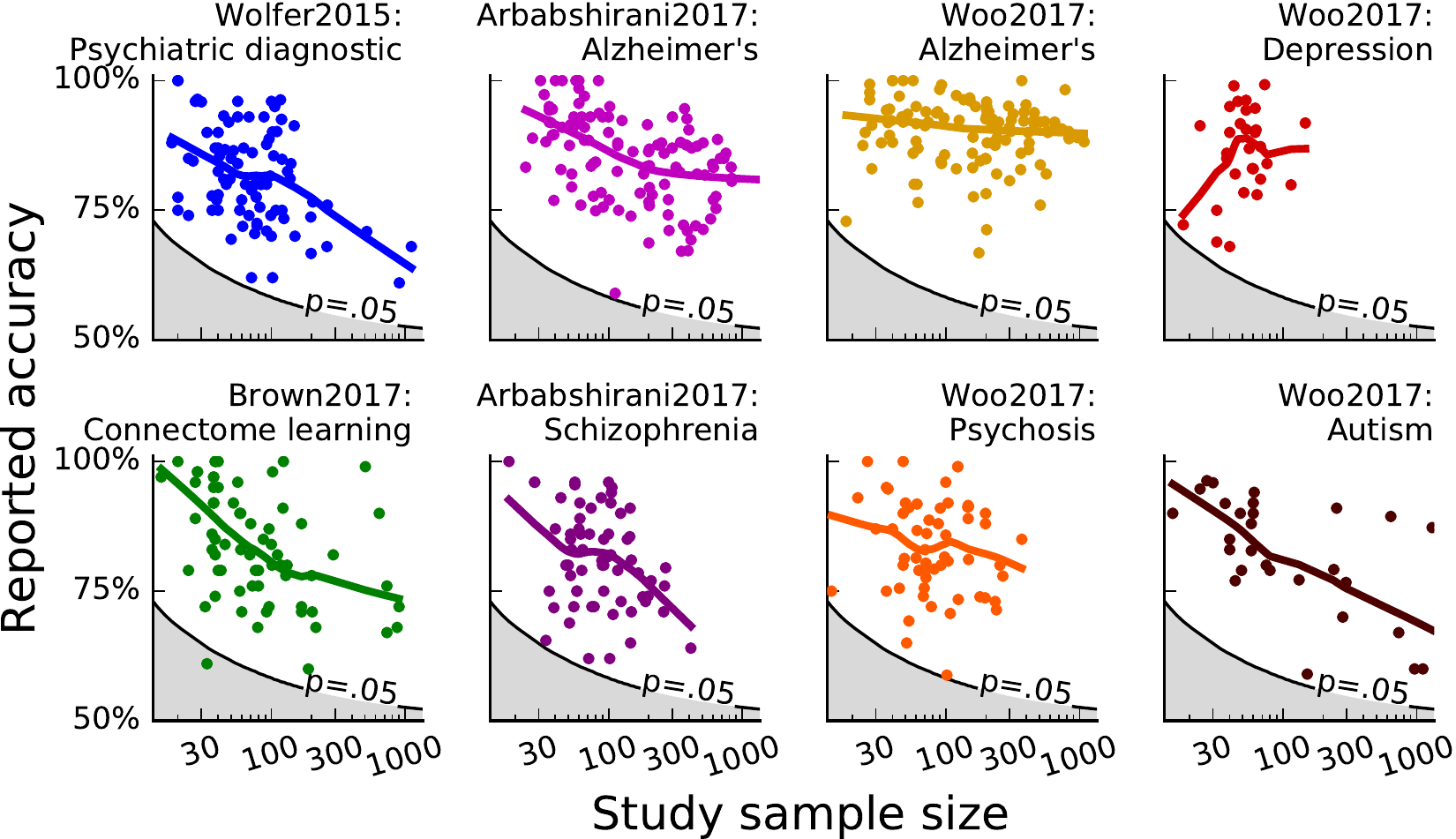}%
    \caption{\textbf{Reported accuracy and sample size}
    The various plots show reported prediction accuracy as a
    function of sample size for the studies in different the
    reviews considered in \autoref{fig:sample_size}.
    The black line and grey area represent the $p=0.05$ threshold with a
    binomial null model, which, as shown in the results section, is
    likely to be optimistic.
    The lines are Lowess fit to the data: robust non-parametric local
    regression.
    \label{fig:reported_accuracy}
    }
\end{figure}

The combination of large variance and the filter effect of publications
could explain why the prediction accuracy reported in publication often
decreases as sample sizes increase. Indeed, in
\autoref{fig:reported_accuracy} I plot an meta analysis uniting the
results discussed in several review papers. Each of these review select a
variety of studies on different criteria such as methodology used or
pathology studied. Overall, the typical prediction accuracy reported in
studies with small samples size is larger that reported in studies with
many samples\footnote{Depression studies, as reported by
\cite{woo2017building} do not show this decrease, however none of these
have a large sample.}. 
Homogeneity of the population and the imaging data is harder to control
on larger cohorts. Hence uncontrolled heterogeneity might explain such a
decrease. However, very few studies have compared large heterogeneous
cohorts to smaller well-controlled group with the same analytic pipeline.
A notable exception, \cite{abraham2017deriving}, finds that pooling data
across sites leads to better predictive biomarkers of Autism, although
this is a highly-heterogeneous spectrum disorder.

\subsection{Cross-validation is nonetheless a crucial tool}
% Defend cross-validation

Cross-validation is not a silver bullet. However, it is the best
tool available, because it is the only non-parametric method to test for
model generalization. Bayesian approaches such as Bayesian model
selection or Bayesian model averaging rely on model evidence to test or
select models \citep[chap.\,35]{penny2007}. However, they are strongly
parametric: the statistical
control or the usefulness of this test collapses if the modeling
assumptions are wrong. Additionally, 
these approaches do not measure the ability of the model to predict on new
data.

Testing for generalization is central to diagnostics or prognosis
applications, where prediction is indeed the question. It has also a
broader importance as the ability to generalize findings is central to
scientific investigations. Research in psychology and neuroscience has
focused on explaining data, to seek causal mechanisms using
tightly-controlled experiments, \emph{eg} based on randomization.
However, too strong a focus on well-controlled explanation may limit the
generality of the results \citep{yarkoni2016choosing}.
The essential aspect of cross-validation is that it tests a model on
observations independent from the data that was used to fit the model.
This is the only assumption-free way to bound model complexity. Indeed,
more complex model will always fit the data better. There are statistical
procedures to set model complexity, such as Bayesian information
criterion (BIC) and the related Akaike information criterion (AIC) and
minimum descriptor length (MDL). However, they rely on modeling
assumption such as data distribution, independence of the observations,
and need much more observations than model parameters
\cite[sec.\,7.5]{hastie2009elements}.

%Note that criticism of CV based on the fact that it leads to
%sub-optimal fits (depletion of the train set) is not related to what we
%are observing here: the error that we are measuring is an error on
%quantifying how well the given model generalizes.

\subsection{Looking forward: some recommendations}

% Be positive
Predictive models can extract richer and finer information from the
complex data provided by brain imaging. However, best practices need to
be adapted to ensure enough statistical power to test these models.
% Acknowledge difficulties
While larger datasets are certainly desirable, they are difficult and
costly to acquire. At the subject level, data accumulation is limited by
fatigue of the subject in the scanner as well as habituation effects to
the paradigm. Scanning many subjects may entails operational budgets
beyond that typical of a neuroimaging grant. Nevertheless, there are a
variety of solutions feasible without major changes in the field.

\paragraph{Data sharing and pooling, despite heterogeneity}
Reusing shared data across investigators can increase sample sizes while
keeping bounds on data-acquisition costs \citep{poldrack2014making}.
Platforms to share neuroimaging data are rapidly growing, as with
OpenfMRI \citep{poldrack2013toward} that now hosts 63 studies comprising
2\,200 subjects, or Neurovault \citep{gorgolewski2015neurovault} with
26\,000 brain maps in 1\,100 collection. Such sharing is easiest with
harmonized protocols and conventions. Yet, outside of concerted efforts,
there is a massive amount of data potentially available: around 30\,000
studies using fMRI are published each year\footnote{As estimated from a
PubMed search on fMRI.}, many with new data. They answer a wide variety
of different questions; still they have some overlap. This overlap
provides opportunity for reuse, increasing sample size. For cognitive
neuroimaging, joint analysis is challenging due to the high specificity
of cognitive questions studied. However, the success of meta-analysis in
fMRI suggests that pooling data can be beneficial, whether it is by
assembling a small number of well-matched studies or over a wider
coverage of the literature \citep{laird2005ale,costafreda2009pooling}.
In a remarkable example of predictive models using pooled data,
\cite{wager2013pain} were able to combine multiple pain studies to
extract a neural signature specific to physical pain, discriminating it
from social pain or warmth.

To pool studies of brain pathologies, it is often easier to define a
common covariate to predict across subjects, typically a diagnostic
status. However, studies of the same pathology can differ in their
inclusion criteria, introducing heterogeneity that confounds predictions
or interpretations. Heterogeneity may be a challenge to the clinical
relevance of studies on heterogeneous groups, as many neuro-psychiatric
diseases are spectrum disorders that are likely composed of several forms
of the disease. 
However, biomarkers that are too specific to a certain site or a certain
cohort have reduced clinical value \citep{woo2017building}. There are
many documented successes of prediction from heterogeneous brain imaging
data. For anatomical markers of aging, \cite{ziegler2014individualized}
show that using data from many scanners enables to generalize to new
scanner. \cite{yahata7small} and \cite{abraham2017deriving} show that,
for a disorder as heterogeneous as Autism, predicting diagnostic status
across sites was possible. Moreover, \cite{abraham2017deriving} and
\cite{dansereau2017statistical} show that with a large number of sites,
prediction across sites performed as well as prediction across subjects
in the same site.
Cross-validation on heterogeneous data requires some care, as prediction
may be driven by a confounding covariate \citep{little2017}. For
instance, when predicting with several sessions per subject, care must be
taken to avoid having different sessions of the same subject in the train
and test set, to prevent subject-identification to be driving prediction
\citep{saeb2017}.

\paragraph{Paradigms facilitating larger data}
Some experimental paradigms make it easier to accumulate data, often to
the cost relinquishing fine control on cognition. For instance, to study
cognition, standard localizer-type paradigms \citep{saxe2006divide} can
easily be shared across many acquisitions, leading to large databases
\citep{pinel2007fast}. Naturalistic stimuli enables faster presentations
for longer times without fatigue of the subject. Therefore they can be
used to accumulate subjects' responses for rich decoding studies
\citep{kay2008identifying}. To study inter-individual differences,
acquisition protocols that are comparatively universal and easy to
acquire lead to large sample sizes. For instance there are more standard
T1 maps available than myelin maps. In functional imaging, resting-state
fMRI acquisition are a promising source of very large data, via post-hoc
aggregation \citep{biswal2010,thompson2014enigma,di2014autism} or large
concerted efforts
\citep{miller2016multimodal,van2013wu}.

\paragraph{Cognitive neuroimaging results: at the group level}
In cognitive neuroimaging, multi-voxel pattern analysis (MVPA) generally
performs cross-validation across trials in the same subject. The number
of trials cannot always be easily extended, due to habituation effects or
limited time in the scanner. A more promising avenue to increase sample
size is to exploit the replication of these decoding results across
subjects. As there is significant variability in cognitive strategy or
performance across subjects, pooling across subjects raises concerns.
Yet, conclusions should be drawn from the group, and not at the subject
level, where the small sample size tends to compromise cross-validation.
There are several approaches. First, as outlined in
\cite{stelzer2013statistical} even when cross-validation is performed at
the subject level, testing for significance of predictions can be done at
the group level. This approach is used by a good fraction of the MVPA
studies. Another option is to predict across subjects. This
requires fine-grain matching of subjects' anatomy and function, yet it
bears the promise of more general representations of cognition
\citep{haxby2011common}.

\paragraph{Evaluating methods on multiple studies}
For methods development, the vibration effects observed on
\autoref{fig:haxby_decoding} are very troublesome. Indeed, the empirical
work in methods development often amounts to trying out multiple
approaches and publishing the one that works best. It leads naturally to
overfit if the data are not large enough to guarantee errors on the
measurement prediction accuracy smaller than the difference between
methods. As I outline in \autoref{sec:underestimated} and
\ref{sec:sem_error}), it is hard to measure these error bars and they are
usually underestimated. The best way to compare approaches without
loophole is to test them across several datasets
\citep{demvsar2006statistical}. With the sample size typical of
neuroimaging, I personally believe that this is the only sound way of
doing methods development. As most methods researchers, I have not always
worked like this in the past, and some of the promising results that we
have published have not carried over\footnote{As an example, we were not
able to reproduce the benefits of the specific algorithm in
\cite{michel2012supervised} on other datasets, though we later validated
some of the core ideas --voxel clustering-- on many other datasets
\cite{varoquaux2012icml,hoyos2016recursive}.}.

\section{Conclusion: improving predictive neuroimaging}

With predictive models even more than with standard statistics small
sample sizes undermine accurate tests. The problem is inherent to the
discriminant nature of the test, measuring only a success or failure per
observations. Estimates of variance across cross-validation folds give a
false sense of security as they strongly underestimates errors on the
prediction accuracy: folds are far from independent. 
Rather, to avoid the illusion of biomarkers that do not generalize or
overly-optimistic methods development, ballpark estimates of confidence
bounds summarized in \autoref{tab:confidence_bounds} may be more useful.
A typical sample size in neuroimaging, 100 observations, leads to
$\pm10\%$ errors in prediction accuracy. Cognitive neuroscience
MVPA studies often control these errors by performing a group-level statistical
analysis.

\begin{table}\center
{\sffamily%
\begin{tabular}{rcccc}
Sample size & 30 & 100 & 300 & 1000
\\
\hline
Confidence bounds & $\pm15\%$  & $\pm10\%$ & $\pm6\%$ & $\pm3\%$
\end{tabular}}
\caption{\textbf{Confidence bounds to be expected for a binary
classification}, summarizing experiments and simulations in
\autoref{fig:cross_val_errors}.
Actual confidence bounds may be significantly larger in adverse
situations such as with correlated observations or very unstable
classifiers.
\label{tab:confidence_bounds}
}
\end{table}

Exploring arbitrary choices in analytic pipelines easily creates
improvements in measured prediction accuracy that will not 
generalize to new data. Such effect is a major impediment for methods
development as it becomes challenging to ensure that improvements
observed are meaningful. Due to the specificities of datasets, protocols,
or pathologies, there cannot be a one-size-fits-all optimal method for
predictive modeling. However, to limit the variety of analytics
pipelines, we, methods developers, must provide general recommendations
validated on many datasets.

With small sample sizes, research with predictive models is performed
blindfolded. The problem is neither new nor specific to neuroimaging. In
genomics, \cite{braga2004cross} have asked ``Is cross-validation valid
for small-sample microarray classification?''. In neuroimaging, it is
magnified by the intrinsic difficulty of acquiring large datasets. 
The problem will not be fixed by better classifiers or cross-validation
approaches. Solutions will
lie in approaches using larger samples sizes or preregistered analyses. Overall, exploring larger
datasets is a promising future for neuroimaging
\citep{poldrack2016scanning}. Their richness is best captured by
multivariate models \citep{miller2016multimodal}.
For predictive applications such as biomarkers, larger datasets lead to
better prediction on hard problems, even in the face of increased
variability.

\subsection*{Acknowledgments}

Computing resources were provided by the NiConnect project
(ANR-11-BINF-0004\_NiConnect). I am grateful to Aaron Schurger, Steve
Smith, and Russell Poldrack for feedback on the manuscript. I would also
like to thank Alexandra Elbakyan for help with the literature review, as
well as Colin Brown and Choong-Wan Woo for sharing data of their review
papers.

%%%%%%%%%%%%%%%%%%%%%%%%%%%%%%%%%%%%%%%%%%%%%%%%%%%%%%%%%%%%%%%%%%%%%%
% references section

\small
\bibliography{biblio}

\appendix
\renewcommand{\thefigure}{A\arabic{figure}}
\setcounter{figure}{0}
\renewcommand{\thetable}{A\arabic{table}}
\setcounter{table}{0}

\section{Additional considerations on uncertainty in prediction accuracy}

\label{sec:additional_considerations}

\subsection{The reusable holdout}

\cite{dwork2015reusable} propose an elegant technique to reuse a given
holdout set while avoiding overfitting it. However, the technique relies
on jittering the measure of prediction error when it is below a
threshold\footnote{Technically, the jitter is performed when train and
test errors are very close to each other. Optimally-tuned predictors
strike a balance between over and under fit and hence have close error
rates on the train and test set.}. The technique does not fix the
intrinsic uncertainty in the measurement of the prediction accuracy --a
task likely impossible-- but it embeds this uncertainty in the validation
procedure, refusing to conclude beyond a threshold directly related to
confidence intervals of the prediction \citep[supp
mat]{dwork2015reusable}. A given control on generalization performance
requires setting the threshold proportional to $\sqrt{n}$. The reusable
holdout is a beautiful improvement to cross-validation, that is however
aligned with the main point that I am making: measuring prediction
accuracy is not reliable with small samples.

\subsection{Confidence bounds for varying expected accuracy}

\begin{figure}[b]
    \centerline{%
      \includegraphics[width=.8\linewidth]{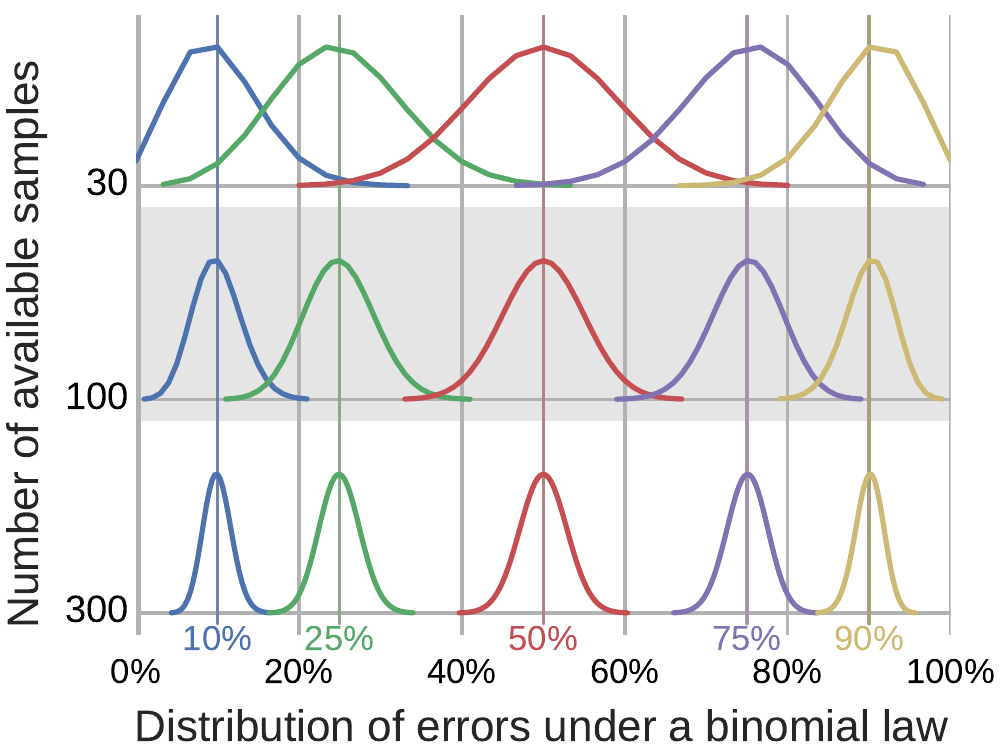}%
    }
    \caption{\textbf{Varying expected accuracy}
    Binomial distributions for varying expected accuracy and number of
    samples. These indicate the shape of sampling noise, whether it is
    for a null distribution or the observed values.
    \label{fig:varying_accuracy}}
\end{figure}

The experiments performed so far are for a chance level of 50\% and an
average prediction accuracy of 75\%. While these numbers are typical in
many decoding experiments, some experiments probe multiclass decoding,
sometimes with many classes, in which case the accuracy under chance as
well as the observed accuracy may be much lower. In such situations, the
mechanisms driving estimation errors in cross-validation are the same,
hence a binomial law still give a lower-bound on the distribution of
errors. The binomial must be adapted to be centered on the expected
accuracy, whether it is to compute the null distribution or to
evaluate confidence bounds on observed values. 
\autoref{fig:varying_accuracy} shows different binomial distributions for 
various values of expected accuracy and number of samples. For
expected accuracy close to 0\% or 100\%, the distributions narrow and
becomes asymmetric due to the censoring effect of these limits. With
large sample sizes, the distributions are more narrow, and these effects
are less visible. \autoref{tab:varying_accuracy} gives corresponding 5
and 95\% confidence bounds and shows that indeed, the confidence bounds
are tighter near 0\% or 100\% prediction accuracy.

\begin{table}
    {\sffamily\small%
    \rowcolors{2}{gray!15}{white}%
    \begin{tabular}{rccc}
    Expected & \multicolumn{3}{c}{5\%--95\% confidence bounds} \\
    accuracy & 30 samples & 100 samples & 300 samples \\
    \cline{2-4}\\[-1.5ex]
    10.0\% & $ 3.3\%$--$20.0\%$ &
             $ 5.0\%$--$15.0\%$ &
             $ 7.3\%$--$13.0\%$ \\
    25.0\% & $13.3\%$--$40.0\%$ &
             $18.0\%$--$32.0\%$ &
             $21.0\%$--$29.0\%$ \\
    50.0\% & $36.7\%$--$63.3\%$ &
             $42.0\%$--$58.0\%$ &
             $45.3\%$--$54.7\%$ \\
    75.0\% & $60.0\%$--$86.7\%$ &
             $68.0\%$--$82.0\%$ &
             $71.0\%$--$79.0\%$ \\
    90.0\% & $80.0\%$--$96.7\%$ &
             $85.0\%$--$95.0\%$ &
             $87.0\%$--$92.7\%$ \\
    \end{tabular}
    }
    \caption{\textbf{Confidence bounds} for a varying expected accuracy
    and varying number of samples, the 5 and 95\% percentile of the
    binomial distribution, giving a lower-bound on the confidence
    bounds as it is a conservative distribution of errors (see \autoref{fig:perfect_predictors}).
    Not that experiments revealed that the binomial distribution
    underestimates errors, hence actual confidence bounds are likely to
    be higher.
    \label{tab:varying_accuracy}}
\end{table}

\section{Experiments of Varoquaux 2017}

\label{sec:varoquaux_2017}

\begin{figure}[h]
\centerline{\includegraphics[width=.7\linewidth]{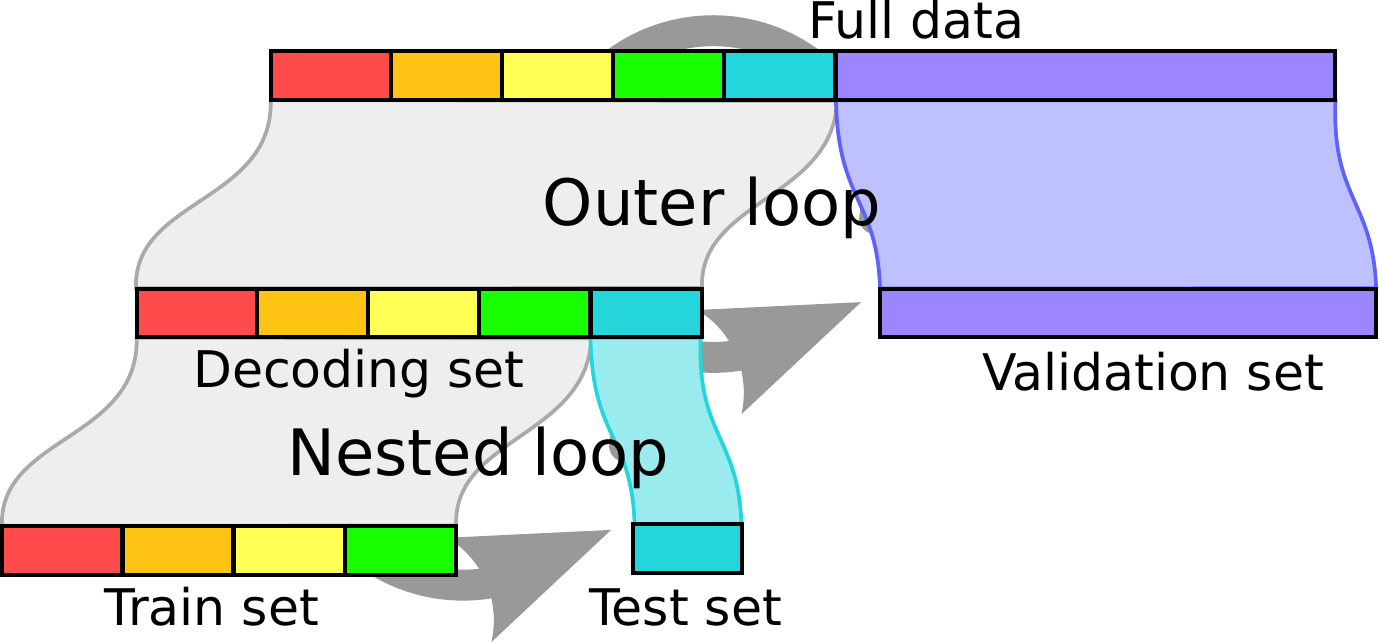}}
\caption{Splitting the data twice: first in validation and decoding set,
and then performing cross-validation on the decoding set.
    \label{fig:nested_cv}
}
\end{figure}

To facilitate reading this paper, I summarize here the experimental
protocol used in \cite{varoquaux2017assessing}. The principle of the
experiment is that the data are split twice (see
\autoref{fig:nested_cv}):
first in a decoding set and a
validation set; then cross-validation is performed on the decoding set
results in an estimate of prediction accuracy --as in any
cross-validation based study--; finally this estimate is compared to the
prediction accuracy of the models on the left-out validation set. To give
a good measure of accuracy on the validation set, this set is taken
large, as large as the decoding set. The estimation error of
cross-validation is then measured by the discrepancy between the
prediction accuracy on the validation set, and the prediction accuracy
obtained by the cross-validation procedure on the decoding set.
\cite{varoquaux2017assessing} applied such experiments on a variety of
neuroimaging decoding datasets, within and across subjects, in fMRI,
VBM (Voxel Based Morphometry) and MEG (Magneto EncephaloGraphy).

\section{Details on the simulations}

% we simulated a normally-distributed
% dataset with controlled separability. This allows us to directly
% attribute the variance in predictive estimates to the varying
% configurations of CV.

\label{sec:simulated_data}

\subsection{Dataset simulation}

\begin{figure}
    \begin{minipage}{.5\linewidth}
    \caption{\textbf{2D view on simulated data}
    The two classes are represented in red and blue  circles. Here, to simplify
    visualization, the data are generated in 2D (2 features), unlike the actual
    experiments, which are performed on 300 features.
    \label{fig:simulated_datasets}}
    \end{minipage}%
    \hfill%
    \begin{minipage}{.45\linewidth}
	\includegraphics[width=\linewidth]{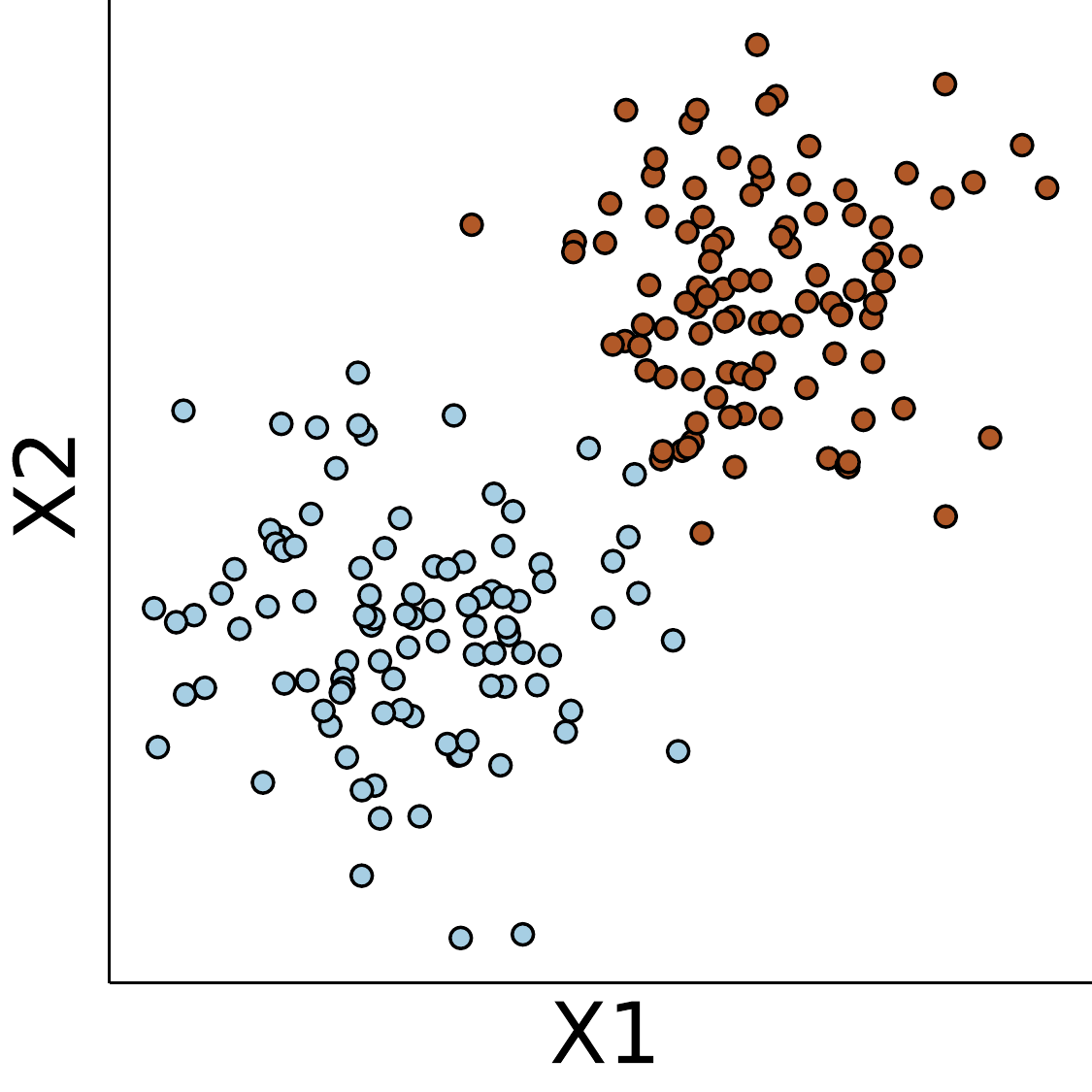}%
    \end{minipage}%
\end{figure}

I generate data with samples from two classes, each described by a
Gaussian of identity covariance in 100 dimensions. The classes are
centered respectively on vectors $(\mu, \dots, \mu)$ and 
$(-\mu, \dots, -\mu)$ where $\mu$ is a parameter adjusted to control the
separability of the classes. With larger $\mu$
the expected predictive accuracy would be higher. The samples are
generated \emph{i.i.d.}, with is a simplification compared to
time-series, as in decoding, where there often is a dependence between
neighboring observations, or in the same session.
I chose the separability $\mu$ empirically to have a
classification accuracy of 75\%. \autoref{fig:simulated_datasets}
shows a 2D view of the corresponding data. 
Code to reproduce the simulations can be found on
\url{https://github.com/GaelVaroquaux/cross_validation_failure}.

\subsection{Experiments on simulated data}
\label{sec:simdataclassifier}

Unlike with a brain imaging datasets, simulations open the door to
measuring the actual prediction performance of a classifier, and
therefore comparing it to the cross-validation measure.

For this purpose, I generate a pseudo-experimental data with a varying
number of train samples, and a separate very large test set, with 
10\,000 samples. The
train samples correspond to the data available during a neuroimaging
experiment, and I perform cross-validation on these. I then apply the
decoder on the test set. The large number of test samples provides a good
measure of prediction power of the decoder \citep{arlot2010}. As a
decoder, I use a linear SVM with C=1, as it is common in neuroimaging. To
accumulate measures, I repeat the whole procedure 1000 times.

\section{Results on the standard error of the mean}

\begin{figure}%
    \center%
    \includegraphics[width=.7\linewidth]{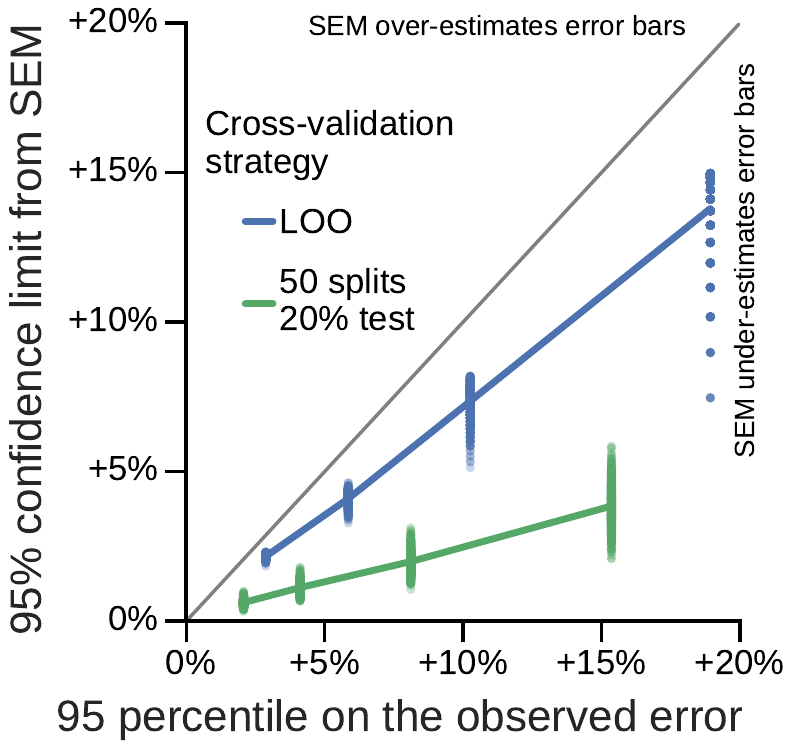}%
    \medskip

    {\small\sffamily%
    \begin{tabular}{llrrr}
    %\toprule
	CV        &       train &          SEM &    empirical \\
	strategy  &        size &    error bar &    error bar \\
    \toprule
    \multirow{4}{9ex}{LOO} & 30 &  $\pm 13.8\%$ &  $\pm 18.9\%$ \\
		&           100 &  $\pm 7.4\%$ &   $\pm 10.3\%$ \\
	        &           300 &  $\pm 4.1\%$ &   $\pm  5.9\%$ \\
		&          1000 &  $\pm 2.2\%$ &   $\pm  2.9\%$ \\
    \midrule                                            
    \multirow{4}{9ex}{50 splits,\\20\% test} & 30 &  $\pm 3.4\%$ & $\pm 15.3\%$ \\
                  &         100 &  $\pm 2.0\%$ &   $\pm  8.1\%$ \\
                  &         300 &  $\pm 1.1\%$ &   $\pm  4.1\%$ \\
	          &        1000 &  $\pm 0.6\%$ &   $\pm  2.1\%$ \\
    %\bottomrule
    \end{tabular}}
    \caption{\textbf{Error bars: SEM estimates versus observed}
    In conventional models, the confidence limits are a factor 1.64 of
    the standard error of the mean (SEM). This figure represents such
    confidence limits on cross-validation estimated from SEM across the
    folds as a function of the actual
    estimation error observed in the simulations. Using the standard
    formula to compute the 95\% confidence limit under-estimates it
    significantly compared to the actual 95 percentile of the observed
    error, thought the two different choices
    of cross-validation strategy, leave one out, and 50-times repeated
    splitting of 20\% of the data, give different  under-estimation: a
    factor of 0.73 for leave one out, and 0.26 for 50 repeat splits.
    \label{fig:error_vs_sem}}
\end{figure}

\label{sec:sem_error}

A common approach to give error bars is to compute the standard error of
the mean (SEM) across the cross-validation folds. For samples drawn from
a normal distribution, the distance from the mean of the upper and lower 
95\% confidence limit is given by 1.64\,SEM \footnote{If the test is
two-sided, the confidence bound are given by 1.96\,SEM.}. The SEM is also the
quantity that appears in a T test. On the simulations, I
compared such confidence limits computed from the SEM to the observed
percentile of \autoref{fig:error_vs_sem}.

Using the standard formula based on the SEM under-estimates actual
confidence bounds by a factor of 0.73 for leave one out and 0.26 for
repeated train-test split with 20\% left out and 50 splits. There is
indeed a wide difference is how much different folds are correlated in a
cross-validation strategy. To give a more precise estimation of
prediction accuracy, repeated random splits create more correlations
across fold, and hence standard SEM computation that ignores this
correlation is more severely incorrect.

\section{Experiments with the perfect predictor}

\begin{figure}
    \centerline{%
     \includegraphics[width=.8\linewidth]{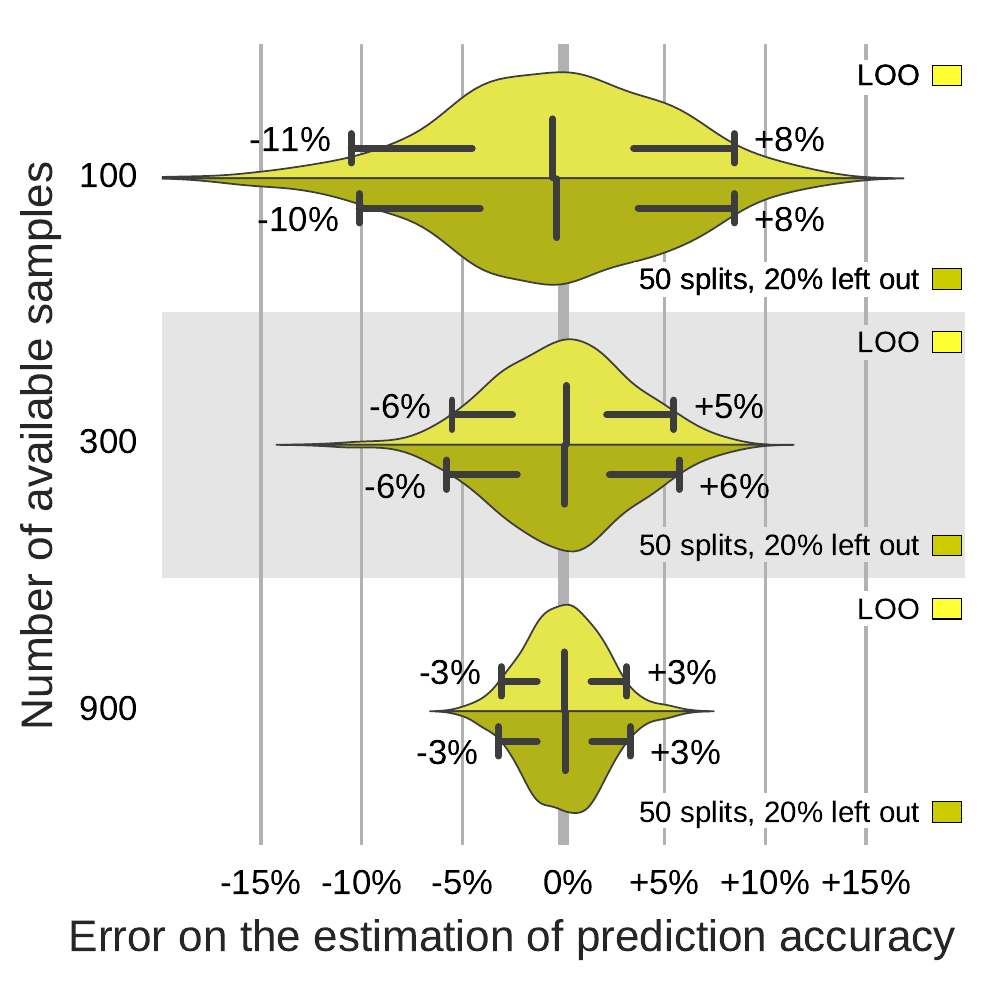}%
     \llap{\raisebox{.72\linewidth}{%
	\parbox{.86\linewidth}{\bfseries\sffamily Perfect\\predictor}}}%
    }%

    \caption{\textbf{Cross-validation error with the perfect predictor.}
    Given a data-independent optimal predictor,
    distribution of errors between the prediction accuracy as assessed
    via cross-validation on data of various sample sizes and the expected
    error of the predictor.
    The bar and whiskers indicate the median
    and the 5\raisebox{.5ex}{\tiny th} and 95\raisebox{.5ex}{\tiny th}
    percentile. The distributions are reported for two reasonable choices
    of cross-validation strategy: leave one out, or 50-times repeated
    splitting of 20\% of the data.
    \label{fig:perfect_predictors}}
\end{figure}

To fully rule out that the errors witnessed on cross-validation are due
to instabilities of the predictive model, I repeated the experiments with
a predictor independent from the data. Specifically, I used the knowledge
of the data-generating process to create a classifier making best
decision possible. I then ran the cross-validation experiments with this
classifier. \autoref{fig:perfect_predictors} gives the corresponding
distribution of mismatch between the accuracy measured by
cross-validation and the actually accuracy of the classifier.

The results with the perfect predictor are very similar to those using an actual decoder trained
on the data\footnote{Note
that I set the separation in the data generation to have a
prediction accuracy of 75\%. As the perfect predictor is a better
predictor than a linear SVC, experiments with the perfect predictor are
done with a large separation.} (\autoref{fig:cross_val_errors}b using a linear SVM).
Given that the classifier is independent of the data, the variability
observed here can clearly be traced to sampling noise in the test set.
Leave-one-out and random splits with 20\% of the data give the
same errors.

%\begin{figure}[t]
%    \centerline{%
%     \includegraphics[width=.8\linewidth]{github_code/dimensionality_results.pdf}%
%     \llap{\raisebox{.72\linewidth}{%
%	\parbox{.86\linewidth}{\bfseries\sffamily Binomial\\law}}}%
%    }%
%
%    \caption{\textbf{Distribution of errors as given by a binomial law}
%    Different between the observation prediction error and the 
%    population value of the probability of error for different sample sizes.
%    The bar and whiskers indicate the median
%    and the 5\raisebox{.5ex}{\tiny th} and 95\raisebox{.5ex}{\tiny th}
%    percentile. 
%    \label{fig:binomial}}
%\end{figure}

\end{document}